\documentstyle[epsfig]{mn2e}
\begin{document}

\title[Cross-correlation of sub-mm and optical galaxies]{On the
  cross-correlation of sub-mm sources and optically-selected galaxies}

\author[Chris Blake et al.]{Chris Blake$^{1,}$\footnotemark, Alexandra
  Pope$^1$, Douglas Scott$^1$, Bahram Mobasher$^2$ \\ \\ $^1$
  Department of Physics \& Astronomy, University of British Columbia,
  6224 Agricultural Road, Vancouver, B.C., V6T 1Z1, Canada \\ $^2$
  Space Telescope Science Institute, 3700 San Martin Drive, Baltimore,
  MD 21218, United States}

\maketitle

\begin{abstract}
Bright sub-mm galaxies are expected to arise in massive highly-biased
haloes, and hence exhibit strong clustering.  We argue that a valuable
tool for measuring these clustering properties is the
cross-correlation of sub-mm galaxies with faint optically-selected
sources.  We analyze populations of SCUBA-detected and optical
galaxies in the GOODS-N survey area.  Using optical/IR
photometric-redshift information, we search for correlations induced
by two separate effects: (1) cosmic magnification of background sub-mm
sources by foreground dark matter haloes traced by optical galaxies at
lower redshifts; and (2) galaxy clustering due to sub-mm and optical
sources tracing the same population of haloes where their redshift
distributions overlap.  Regarding cosmic magnification, we find no
detectable correlation.  Our null result is consistent with a
theoretical model for the cosmic magnification, and we show that a
dramatic increase in the number of sub-mm sources will be required to
measure the effect reliably.  Regarding clustering, we find evidence
at the $3.5$-$\sigma$ level for a cross-correlation between sub-mm and
optical galaxies analyzed in identical photometric redshift slices.
The data hint that the sub-mm sources have an enhanced bias parameter
compared to the optically-selected population (with a significance of
2-$\sigma$).  The next generation of deep sub-mm surveys can
potentially perform an accurate measurement of each of these
cross-correlations, adding a new set of diagnostics for understanding
the development of massive structure in the Universe.
\end{abstract}

\section{Introduction}
\renewcommand{\thefootnote}{\fnsymbol{footnote}}
\setcounter{footnote}{1}
\footnotetext{E-mail: cab@astro.ubc.ca}

Deep sub-mm surveys, using the Submillimetre Common User Bolometer
Array (SCUBA; Holland et al.\ 1999) at the James Clerk Maxwell
Telescope, have revolutionized our understanding of the high-redshift
Universe by discovering a new population of distant highly
star-forming dusty galaxies (see Blain et al.\ 2002 and references
therein).  Despite the roughly 15 arcsec SCUBA beam-size (at 850
$\mu$m) and their typical optical faintness, these sources have
gradually been optically identified, aided by a combination of deep
radio observations from the Very Large Array (e.g.\ Ivison et
al.\ 2002) and, more recently, data from the {\sl Spitzer Space
  Telescope\/}.  Follow-up spectroscopy of the counterparts has
confirmed that the median redshift of the sub-mm population is high,
$z \sim 2$ (Chapman et al.\ 2005).

As the depth and area of sub-mm surveys increase, the population can
be characterized statistically by determining basic properties such as
the luminosity function and the clustering amplitude.  Such
measurements will allow the sub-mm population to be described within
the framework of models for the formation of massive galaxies
(e.g.\ Baugh et al.\ 2005).  The clustering properties of a class of
galaxies, interpreted in terms of cold dark matter type models, is a
key measurement (see e.g.\ van Kampen et al.\ 2005).  The bias of a
galaxy population traces the global environment it inhabits, and can
be linked to a representative mass of dark matter halo.  Such
information reflects upon the formation mechanism of the population,
and allows evolutionary sequences to be inferred between objects at
high and low redshifts.  In particular, if sub-mm galaxies originate
from galaxy mergers, then they should be highly biased with respect to
the underlying dark matter, given that mergers occur in high-density
environments in the high-redshift Universe (Percival et al.\ 2003).
If the bias is found to be high, then this would constitute direct
evidence that sub-mm galaxies are indeed the progenitors of today's
massive elliptical galaxies.

Existing surveys of sub-mm galaxies are inadequate for accurately
performing a direct determination of their clustering properties via
measurement of auto-correlation functions.  So far there have been at
best tentative detections of such clustering (e.g.\ Blain et
al.\ 2004).  An alternative approch is to measure instead the
cross-correlation function of the sub-mm population and the dark
matter distribution traced by more numerous optically-selected
galaxies.  Such an analysis was performed by Almaini et al.\ (2005),
using the 8-mJy SCUBA surveys of Scott et al.\ (2002) and the
shallower scan map of the Hubble Deep Field (Borys et al.\ 2002).  A
significant cross-correlation was detected between these sub-mm
data-sets and optical follow-up images.  Interestingly, the measured
correlation was between the sub-mm sources at high redshift and
relatively bright optical galaxies at lower redshifts.  This led
Almaini et al.\ to suggest that the most likely explanation for the
cross-correlation was the phenomemon of cosmic magnification, by which
the background sub-mm galaxies experience gravitational lensing by
foreground dark matter haloes traced by the bright optically-selected
galaxies.  A similar effect has recently been observed for the
cross-correlation between background quasars and foreground galaxies
in the Sloan Digital Sky Survey (Scranton et al.\ 2005).

The amplitude of the cosmic magnification is determined by several
factors, including the dark matter power spectrum and growth function,
but also the slope of the flux distribution of the background
population.  In the case of sub-mm galaxies this slope is
exceptionally steep, leading to a relatively large cross-correlation.
A similar explanation was posited to explain an earlier measured
correlation between sub-mm galaxies and X-ray selected sources
(Almaini et al.\ 2003).  An alternative possibility was also
suggested: that a higher-than-expected fraction of sub-mm galaxies lie
at relatively low redshifts $z \la 1$, such that the two classes of
objects partially trace the same large-scale structure.

In this study we analyze the cross-correlation between sub-mm and
optically-selected galaxies using data-sets from the Great
Observatories Origins Deep Survey North (GOODS-N) region (Giavalisco
et al.\ 2004).  There are several advantages to using this field: (1)
a robust and substantial catalogue of sub-mm sources exists, extracted
from a well-understood compilation of SCUBA data (Borys et al.\ 2003,
2004; Pope et al.\ 2005); (2) there is almost complete identification
of the SCUBA counterparts, including spectroscopic redshifts for
almost half of the objects and photometric redshift estimates for the
rest; and (3) deep {\sl Hubble Space Telescope (HST)\/} ACS
observations have been taken of the entire field, providing a high
density of optical galaxies with photo-$z$ estimates.  The data-sets
used are described in more detail in Section \ref{secdata}.  We pay
particular attention to the estimator for the cross-correlation
function employed, as explained in Section \ref{secest}.  Using
different photometric redshift cuts, we search for cross-correlations
induced by cosmic magnification and by galaxy clustering.  Our
measurements are detailed in Section \ref{secmeas}, and are compared
with theoretical predictions in Section \ref{secmod}.  Prospects for
future sub-mm surveys are considered in Section \ref{secfutsurv}.

\begin{figure}
\center
\epsfig{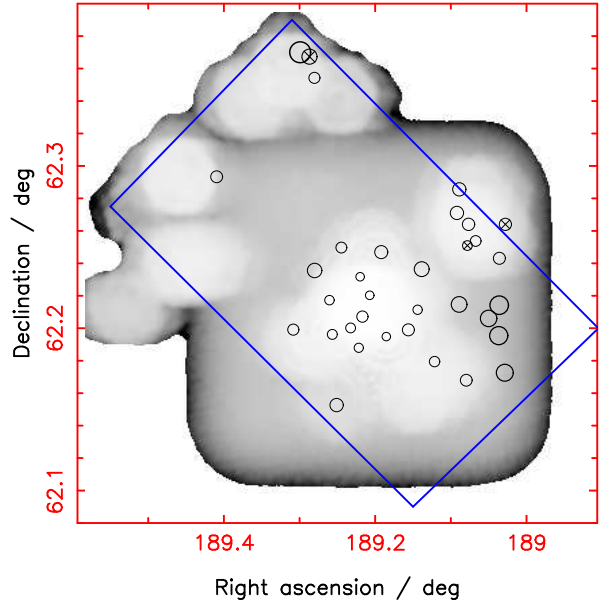}
\caption{Noise map for SCUBA observations in the GOODS-N field.  The
  grey-scale represents the noise level, ranging from a minimum of
  $0.3$ mJy beam$^{-1}$ (white) to a maximum of $15$ mJy beam$^{-1}$
  (black).  The circles indicate the positions of the secure sample of
  34 counterparts extracted from the sub-mm map, with the size of the
  plotted symbol increasing with the brightness of the source.  We
  note that the noise at the positions of the extracted sources ranges
  from $0.3$ -- $4.3$ mJy beam$^{-1}$.  The straight boundaries
  illustrate the extent of the optical ACS observations, which have
  almost uniform sensitivity within this region.  Three of the 34
  sub-mm sources, marked with crosses, are excluded from our
  cross-correlation analysis.  One lies just outside the area of
  uniform optical sensitivity, while two others are within 30 arcsec
  of higher signal-to-noise sub-mm sources where the object extraction
  is particularly difficult, as discussed in Section \ref{secrand}.}
\label{figfield}
\end{figure}

\section{Data-sets}
\label{secdata}

The GOODS-N region covers an area of approximately $10 \times 16.5$
arcmin, centred on $12^{\rm{h}} 36^{\rm{m}} 55^{\rm{s}}$, $+62^{\circ}
14\arcmin 15\arcsec$ (Giavalisco et al.\ 2004).  All of the SCUBA data
from several extensive imaging campaigns in the GOODS-N field have
been combined into one sub-mm map, which we refer to as the
`super-map' (see Borys et al.\ 2003 and references therein).  The
resulting noise map and positions of extracted sources are displayed
in Fig.~\ref{figfield}.  Since the super-map is a compilation of
essentially all SCUBA data taken in the field, the associated noise
map is extremely non-uniform.  The most recent version of the GOODS-N
super-map contains forty 850-$\mu$m sources detected above
3.5-$\sigma$ (Pope et al.\ 2005).  From Monte Carlo simulations, we
expect to find $\sim 3$ spurious sources in the extraction process
(Borys et al.\ 2003).  In order to refine our secure catalogue, we
have explored the level of flux boosting (also referred to as
Malmquist and/or Eddington bias) of these sources by applying the
Bayesian approach discussed in Coppin et al.\ (2005).  By simulating a
distribution of pixel values, which depends on the chopping pattern,
we determined the de-boosted flux for each of our sources.  We then
removed any sources from our sample which were severely affected by
flux-boosting and possessed a non-negligible probability of having
zero flux, which left us with 35 secure sub-mm sources.  Note that the
decision to reject sources due to likely flux boosting comes from a
simple relationship based on signal and noise, and therefore it is
easy to include the same criterion in our simulated sub-mm catalogues
(discussed in Section \ref{secrand}).

Using the optical, radio and new {\sl Spitzer} data in GOODS-N we have
identified counterparts for all but one of the secure sub-mm sources
(the details are discussed in Pope et al.\ 2006).  The positions of
the remaining 34 are plotted in Fig.~\ref{figfield}.  Spectroscopic
redshifts are known for about half of the sub-mm catalogue; reliable
photometric redshifts have been estimated for the remainder of the
objects using the extensive optical and infra-red data (see Figure 1
of Pope et al.\ 2005 and Figure 7 of Pope et al.\ 2006).  For those
sub-mm sources possessing both spectroscopic and photometric
redshifts, the standard deviation of $(z_{\rm phot} - z_{\rm spec})/(1
+ z_{\rm spec})$ is $0.10$.

The optical data for the GOODS-N region (Giavalisco et al.\ 2004),
obtained with the Advanced Camera for Surveys (ACS) on-board the {\sl
  HST}, has a uniform sensitivity within the boundaries indicated in
Fig.~\ref{figfield}, which represents the area studied in our
cross-correlation analysis.  We restricted our analysis to sources
which are detected above 5-$\sigma$ and therefore, for all but the
faintest magnitudes, we are far enough above the noise level that the
subtle variations in exposure time will not affect our results.  In
addition to the 4 ACS bands ($B_{\rm{435}}$, $V_{\rm{606}}$,
$i_{\rm{775}}$ and $z_{\rm{850}}$), the GOODS-N field has also been
surveyed with several ground-based facilities, providing data in the
following bands: \emph{U} (KPNO, Capak et al.\ 2004); \emph{B, V, R,
  I, z} (Subaru, Capak et al.\ 2004); and $J$, $K_{\rm{s}}$ (KPNO,
Giavalisco et al.\ 2004).  Using all of these optical data,
photometric redshifts have been calculated for roughly half of the
$\simeq 32{,}000$ optically-detected galaxies in GOODS-N.  The
accuracy of these photometric redshifts has been determined using the
subset of sources (numbering $\simeq 1{,}700$) which also possess
spectroscopic redshifts.  The standard deviation of $(z_{\rm phot} -
z_{\rm spec})/(1 + z_{\rm spec})$ for the optical catalogue is 0.11.
We note that throughout this paper we use AB magnitudes.

\section{The cross-correlation function}
\label{secest}

The cross-correlation function $w_{\rm cross}(\theta)$ between two
galaxy populations 1 and 2, in terms of an angular scale $\theta$, is
defined as the fractional excess in the probability $\delta P$,
relative to a random unclustered distribution, of finding both a
galaxy of type 1 in a solid angle element $\delta \Omega_1$ and a
galaxy of type 2 in a solid angle element $\delta \Omega_2$, separated
by angle $\theta$:
\begin{equation}
\delta P = \Sigma_1 \, \Sigma_2 \, [1 + w_{\rm cross}(\theta) ] \,
\delta \Omega_1 \, \delta \Omega_2,
\end{equation}
where $\Sigma_1$ and $\Sigma_2$ are the surface densities of
populations 1 and 2 (Peebles 1980).

The cross-correlation function $w_{\rm cross}(\theta)$ is measured
from the galaxy distributions by constructing {\it pair counts} from
the data-sets.  A pair count between two galaxy populations 1 and 2,
$D_1D_2(\theta)$, is a binned histogram of the separations $\theta$ of
every galaxy of population 1 relative to all objects of population 2.
In order to determine the cross-correlation function, fully
incorporating the effects of the survey geometry and of statistical
fluctuations, we must also generate random unclustered realizations of
the galaxy distributions with the same angular selection functions as
the real data.  The pair counts between the data and random
distributions (denoted $D_iR_j$) measure the actual average available
area around each object, taking account of the survey window function
and the distributions of the galaxies relative to the boundaries of
the sample.  In this way, we can construct correlation function
estimators with the smallest bias and variance in the angular range
under investigation.

The error in a correlation function estimator is determined from the
variance of individual pair counts.  It is important to note that, if
a separation bin contains $N$ galaxy pairs, then the statistical
variance in this bin will in general exceed the `Poisson error'
$\sqrt{N}$, even for an unclustered distribution of objects, as can be
demonstrated by simulations or analytic calculations (e.g.\ Landy \&
Szalay 1993; Hamilton 1993; Bernstein 1994; Bernardeau et al.\ 2002).
The increase in variance compared to the Poisson prediction depends on
the survey geometry, but can be considerable for a sub-optimal
estimator when the pair separation $\theta$ is not negligible compared
to the survey dimensions.  A fundamental cause of the excess
non-Poisson variance is {\it edge effects}: the position of sources
relative to the boundaries of the sample is important in determining
the distribution of pair separations (i.e.\ a source distant
$\theta_0$ from an edge is less likely to participate in pairs of
separation $\theta > \theta_0$).  The true variance of the correlation
function estimator may be measured by techniques such as jack-knife
re-sampling or Monte-Carlo simulations.

Various estimators for the cross-correlation function have been
proposed.  Two commonly used but (in general) sub-optimal estimators
are:
\begin{equation}
w_{\rm cross}(\theta) = \frac{D_1D_2}{D_2R_1} - 1 ;
\label{eqbadest1}
\end{equation}
\begin{equation}
{\rm and}\ w_{\rm cross}(\theta) = \frac{D_1D_2}{D_1R_2} - 1 .
\label{eqbadest2}
\end{equation}
These two estimators are potentially biased, because in each case
random realizations of only one of the two data-sets have been
created, thus the statistical fluctuations and edge effects due to the
distribution of sources in the other data-set have not been taken into
account.  Furthermore, equations \ref{eqbadest1} and \ref{eqbadest2}
are not invariant when the indices $(1,2)$ are
interchanged.\footnote{This is a clue hinting that the estimators are
  biased, since linear bias terms in one or other data-set may still
  be present.  In general a good cross-correlation estimator should
  not answer {\it one \/} of the 2 questions `is data-set 1 correlated
  with data-set 2' {\it or \/} `is data-set 2 correlated with data-set
  1', but should answer both.}  In addition the variance of these
estimators may significantly exceed the Poisson prediction, depending
on the survey geometry (e.g.\ Landy \& Szalay 1993).  Better
estimators for the cross-correlation function are:
\begin{equation}
w_{\rm cross}(\theta) = \frac{D_1D_2 \times R_1R_2}{D_1R_2 \times
  D_2R_1} - 1 ;
\label{eqgoodest1}
\end{equation}
\begin{equation}
{\rm and}\ w_{\rm cross}(\theta) = \frac{D_1D_2 - D_1R_2 - D_2R_1 +
  R_1R_2}{R_1R_2} .
\label{eqgoodest2}
\end{equation}
These two estimators are modified versions of those originally
suggested for the auto-correlation function by Hamilton (1993) and
Landy \& Szalay (1993), respectively.  In each case statistical
fluctuations in both data-sets have been incorporated, and the
equations are symmetrical in the indices $(1,2)$.

\section{Cross-correlation measurements}
\label{secmeas}

\subsection{Generation of random catalogues}
\label{secrand}

In order to measure the cross-correlation function robustly we must
generate random (unclustered) comparison data-sets for each of our
surveys, possessing the same angular selection functions as the survey
data.  For the optical observations, we generated random catalogues by
uniformly populating the region delineated by the straight boundaries
in Fig.~\ref{figfield}.  One of the 34 secure sub-mm counterparts lies
outside the area of uniform optical sensitivity, as indicated in
Fig.~\ref{figfield}, and as a result was excluded from our analysis.

For the sub-mm data-set, we generated random distributions using the
noise map plotted in Fig.~\ref{figfield}.  Firstly, candidate sources
were randomly generated inside the analysis region from the flux
distribution fitted to this sub-mm data-set by Borys et al.\ (2003):
\begin{equation}
N(S) \propto \left[ \left( \frac{S}{S_0} \right)^\alpha + \left(
  \frac{S}{S_0} \right)^\beta \right]^{-1} ,
\label{eqfluxdist}
\end{equation}
where $N(S) \, dS$ is the number of sources with fluxes between $S$
and $S + dS$.  The best-fitting values of the parameters are
approximately $\alpha = 1$, $\beta = 3.3$, and $S_0 = 1.8$ mJy.  This
distribution represents a number-counts slope steepening with
increasing flux, and is a reasonable fit to all existing SCUBA data
(although the detailed form of this function is not important -- any
function which fits the current data would suffice for our purpose).
The candidate source was only retained if its signal-to-noise ratio
determined by the noise map exceeded $3.5$ and if it satisfied the
flux-deboosting criterion described in Section \ref{secdata}.

Secondly, we must take account of the angular resolution of the sub-mm
observations, otherwise the highly non-uniform noise distribution will
cause the deepest portions of the map to be over-populated by random
sources in comparison to the real data.  Although the beam size of the
SCUBA instrument at $850\,\mu$m is about 15 arcsec, the sub-mm
catalogue displays a dearth of pairs separated by less than 30 arcsec
(compared to the number of pairs expected by random chance), owing to
the difficulty of fitting very close pairs of sources in the
extraction process.  Therefore, we rejected a candidate random source
if its putative position was closer than 30 arcsec to an existing
random source with a higher signal-to-noise ratio.  If the near
neighbour possessed a lower signal-to-noise ratio than the candidate
object, it was expunged from the random catalogue in favour of the new
source.  The final distributions of fluxes in our random catalogues
were found to agree reasonably well with that observed for the real
sub-mm data-set.  The sub-mm catalogue plotted in Fig.~\ref{figfield}
does in fact contain two pairs separated by less than 30 arcsec: for
consistency, the source with the lowest signal-to-noise ratio was
removed in each case, leaving us with a total of 31 sub-mm objects.

\subsection{Determination of errors}
\label{secerr}

As discussed in Section \ref{secest}, the assumption of Poisson errors
can be a poor approximation for the variance in a correlation function
measurement.  Moreover, this approach cannot establish the covariances
between separation bins.  A common technique for improving the error
determination is jack-knife re-sampling, in which correlation
functions are measured for many sub-samples of the data-sets in order
to estimate the statistical fluctuations and covariance
(e.g.\ Scranton et al.\ 2002).  However, our sub-mm catalogue is too
small to allow the reliable application of this method.

We therefore determined the covariance matrix of each correlation
function measurement using Monte Carlo realizations, in which the
actual data-sets were substituted by random realizations generated as
described in Section \ref{secrand}.  This is an acceptable
approximation given that the angular clustering of these data-sets is
weak compared to the shot noise error.  Each mock realization of the
data was analyzed by our correlation function pipeline, and the
results for many realizations enabled the covariance matrix to be
constructed.

\subsection{Bias and variance of estimators}

\begin{figure}
\center
\epsfig{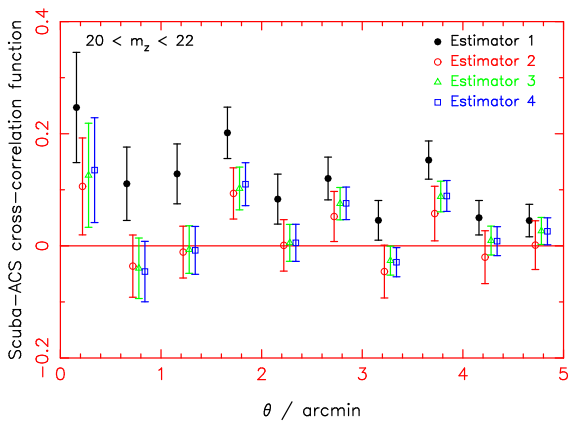}
\caption{The cross-correlation function of the SCUBA and ACS data-sets
  measured by a variety of estimators.  In this test case we
  restricted the ACS sample to galaxies in the $z$-filter magnitude
  range $20 < m_z < 22$.  Estimators 1 to 4 correspond to equations
  \ref{eqbadest1}, \ref{eqbadest2}, \ref{eqgoodest1} and
  \ref{eqgoodest2}, respectively, whilst labels `1' and `2' in these
  equations refer to the optical and sub-mm data-sets, respectively.
  The errors are determined by Monte Carlo realizations.  The separate
  measurements are offset along the $x$-axis for clarity.}
\label{figest}
\end{figure}

Fig.~\ref{figest} indicates how the measured cross-correlation
function depends on the estimator employed, analyzing a test case
consisting of all 31 sub-mm sources and a bright $z$-filter magnitude
slice of ACS galaxies ($20 < m_z < 22$).  Throughout this paper we
measure the cross-correlation function up to an angular scale of 5
arcmin in 10 bins of width $0.5$ arcmin.  The plotted error bars
always correspond to the diagonal elements of the covariance matrix
measured as described in Section \ref{secerr}.  The `good' estimators
for $w_{\rm cross}(\theta)$ (equations \ref{eqgoodest1} and
\ref{eqgoodest2}) produce mutually consistent results that display no
evidence for cross-correlation.  `Estimator 1' (equation
\ref{eqbadest1}) displays a strong positive cross-correlation.
`Estimator 2' (equation \ref{eqbadest2}) is unbiased for these
distributions, but possesses a significantly increased variance.

For `estimator 1', which displays the strong bias in
Fig.~\ref{figest}, we averaged over random optical data-sets, but not
random sub-mm catalogues.  This is problematic, because our actual
sub-mm data-set (Fig.~\ref{figfield}) happens by chance to have more
sources in the lower half of the field than the average realization,
accidently coinciding with an overdensity in the optical data-set.  In
terms of pair counts, $D_{\rm sub-mm}D_{\rm opt}$ is spuriously high
in comparison to $D_{\rm sub-mm}R_{\rm opt}$, causing the positive
offset in the cross-correlation function over a range of scales.  We
emphasize that this is purely an artefact of using a sub-optimal
estimator -- these large-scale fluctuations are entirely consistent
with random realizations of the data-sets, but poor estimators can
mistakenly assign significance to this.

\begin{figure}
\center
\epsfig{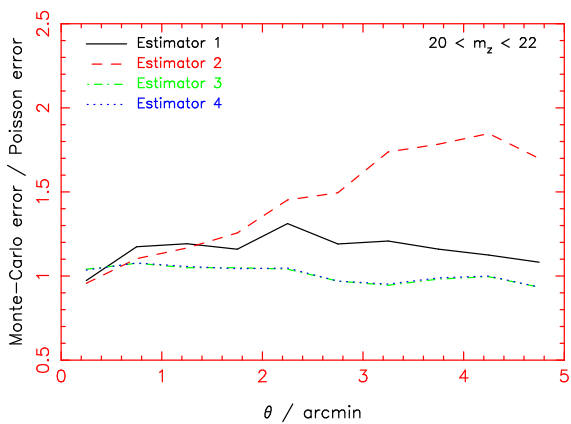}
\caption{The standard deviation in the cross-correlation function of
  the SCUBA and ACS data-sets measured by a variety of estimators, as
  determined by Monte Carlo realizations.  In this test case we
  analyzed the same sample of ACS galaxies as in Fig.~\ref{figest},
  and estimators 1 to 4 have the same correspondences.  The errors are
  normalized to the prediction for purely Poisson statistics.}
\label{figerr}
\end{figure}

In Fig.~\ref{figerr} we plot the ratio between the correlation
function error determined by the Monte Carlo realizations and the
Poisson error, as a function of scale for the four estimators.  The
`good' estimators of equations \ref{eqgoodest1} and \ref{eqgoodest2}
perform best in terms of variance as well as bias, approaching the
Poisson prediction.  `Estimator 2' (equation \ref{eqbadest2})
possesses a significantly larger variance than that predicted by
Poisson statistics.

For the rest of this paper we chose to use the Landy-Szalay-based
estimator for the cross-correlation function (equation
\ref{eqgoodest2}, `estimator 4').  In all cases we determined the pair
counts $D_1R_2$, $D_2R_1$ and $R_1R_2$ by averaging over 10 random
catalogues, each containing the same number of galaxies as the real
survey data-sets.

\subsection{Attempted detection of cosmic magnification}
\label{secmagmeas}

One potential source of genuine cross-correlation between our
data-sets is gravitational lensing (`cosmic magnification') of
background sub-mm sources by dark matter haloes traced by foreground
optical galaxies; we investigate this effect first.  In order to
optimize any detection of cosmic magnification, we used the
photometric redshift information to restrict our comparison to
low-redshift ($0.2 < z < 0.8$) optical sources and high-redshift ($1 <
z < 5$) sub-mm objects.  We only utilized galaxies with `high-quality'
photometric redshifts with errors better than $\delta z = 0.4$ ($95\%$
confidence limit, see Mobasher et al.\ 2004).

\begin{figure}
\center
\epsfig{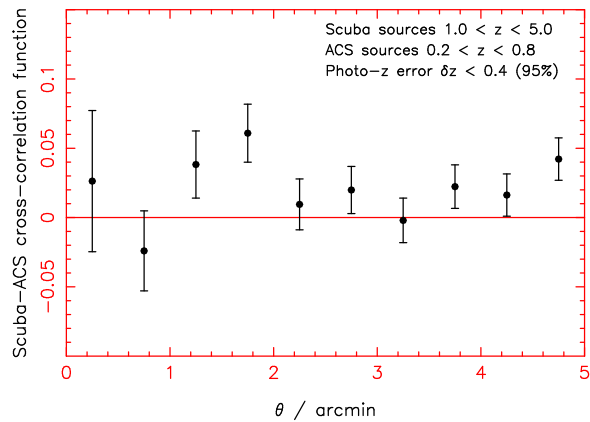}
\caption{Cross-correlation function of the SCUBA and ACS data-sets,
  measured for samples selected in a manner which should be efficient
  for the detection of cosmic magnification.  The measurements contain
  weak evidence for a small positive constant offset from zero, but we
  do not interpret this as an astrophysically-significant
  cross-correlation, as explained in the text.}
\label{figcosmag}
\end{figure}

The result is plotted in Fig.~\ref{figcosmag}.  There is weak evidence
that $w_{\rm cross}(\theta)$ is inconsistent with zero.  In fact a
small positive constant value $w \approx 0.02$ provides a good fit
(using the full covariance matrix).  However, we do not interpret this
as evidence for an astrophysically-significant cross-correlation,
given that the expected form of such a correlation is strongly
scale-dependent (see Section \ref{secmod}).  In contrast, any residual
systematic problems -- such as uncertainties in the underlying mean
source density, the integral constraint correction for small fields,
or unrecognised calibration fluctuations -- would produce a small
constant offset in the correlation function (e.g.\ Blake \& Wall
2002).

\begin{figure}
\center
\epsfig{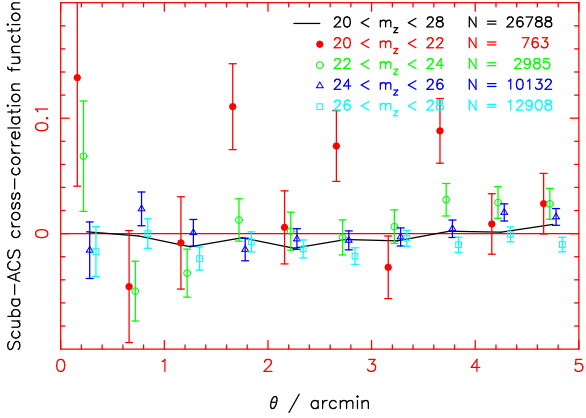}
\caption{Cross-correlation function of the SCUBA and ACS data-sets,
  measured in $z$-filter magnitude intervals of the ACS sources.  The
  solid line is the cross-correlation result using the whole optical
  sample.  The separate measurements are offset along the $x$-axis for
  clarity.}
\label{figmagbin}
\end{figure}

\begin{figure}
\center
\epsfig{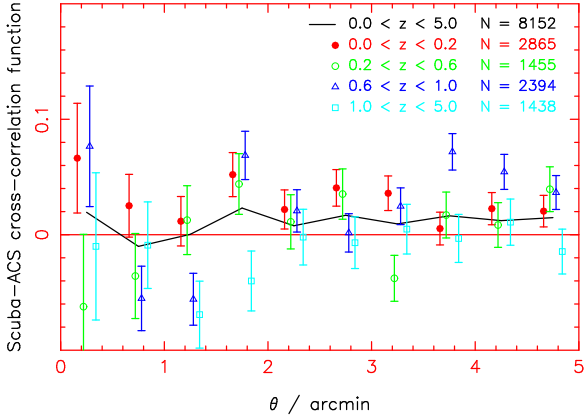}
\caption{Cross-correlation function of the SCUBA and ACS data-sets,
  measured in redshift bands of the ACS sources.  The solid line is
  the cross-correlation result using the whole optical sample.  Only
  optical galaxies with `high-quality' photometric redshifts are
  included in the analysis.  The separate measurements are offset
  along the $x$-axis for clarity.}
\label{figzbin}
\end{figure}

Exploring further, we also measured the cross-correlation function
between all SCUBA sources and various sub-samples of the ACS galaxies.
Fig.~\ref{figmagbin} displays the results measured in different
magnitude bands.  Fig.~\ref{figzbin} plots the measurements in
redshift bands, only including those ACS galaxies with high-quality
photometric redshifts.  In no case do we find a significant detection
of a cross-correlation (that cannot be fit by a small constant
offset).  We also detected no significant cross-correlation by varying
the flux threshold of the sub-mm galaxies.  Bright ($S_{850 \mu {\rm
    m}} > 10$ mJy) sub-mm sources are expected to have the steepest
flux distribution and hence the strongest cross-correlation.  However,
the number of such sources in our sample is very small.

\subsection{Attempted detection of galaxy clustering}

The second potential source of genuine cross-correlation between our
data-sets is galaxy clustering, which would arise if the sub-mm and
optical objects traced the same population of dark matter haloes in
the case of overlapping redshift distributions.  In order to
efficiently search for clustering we again used the photometric
redshift information, firstly restricting our data-sets to galaxies
with high-quality photometric redshifts as defined above.  We also
limited the input sub-mm and optical data-sets to the redshift range
$0.8 < z < 2.0$ in order to maximize the overlap of the catalogues.
We then measured the pair counts in angular separation bins, weighting
each galaxy pair by a factor depending on its redshift difference
$\delta z = z_1 - z_2$:
\begin{equation}
{\rm Weight} = \exp{ \left[ -\frac{1}{2} \left( \frac{\delta
      z}{C} \right)^2 \right] } ,
\label{eqweight}
\end{equation}
where $C = 0.1$ (determined empirically to optimize the
signal-to-noise ratio of the measurement).  These weights strongly
increase the contribution of pairs at similar redshifts, optimizing
any detection of mutual clustering.  The cross-correlation function
was then determined using the usual combination of the (weighted) pair
counts.  The redshifts of sources in the random comparison catalogues
were assigned by randomizing the redshifts of the real data.  The
error in the correlation function was determined using Monte Carlo
realizations, as before.

\begin{figure}
\center
\epsfig{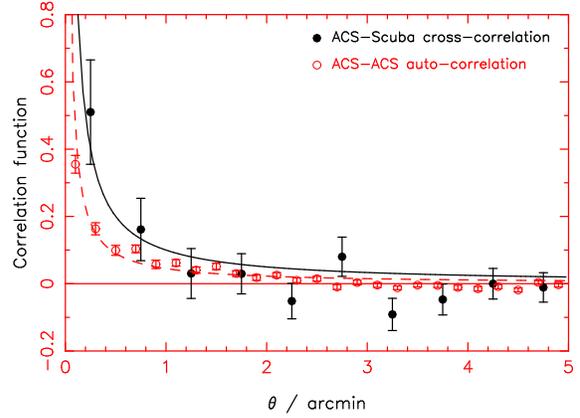}
\caption{The solid circles plot the cross-correlation function of the
  SCUBA and ACS data-sets, measured in a manner designed to optimize
  any detection of galaxy clustering.  The open circles display a
  measurement of the same statistic for an auto-correlation function
  of the ACS galaxies.  The lines represent the best fits of a
  power-law model $w(\theta) = A \, \theta^{-1}$ to separations in the
  range $\theta < 3$ arcmin.}
\label{figclus}
\end{figure}

The result is plotted as the solid circles in Fig.~\ref{figclus}, and
shows evidence for a positive correlation on small scales.  Fitting a
power-law model $w(\theta) = A \, \theta^{-1}$ to the result allows us
to reject a model with no correlation ($A = 0$) with a significance
level of $3.5$-$\sigma$.  The slope of $1.0$ was determined by fitting
a power law to the higher signal-to-noise auto-correlation function of
the ACS galaxies, as described below, and provides a better fit than
the canonical slope of $0.8$.  We checked that adding data with
lower-quality photometric redshifts or changing the value of $C$ in
equation \ref{eqweight} did not improve the significance of the
detection.

For comparison, we repeated the calculation analyzing the clustering
of the ACS galaxies alone via an auto-correlation function, weighting
galaxy pairs as before.  The result, plotted as the open circles in
Fig.~\ref{figclus}, can be established with greater significance,
owing to the larger optical sample.  The auto-correlation amplitude
appears to be lower than the cross-correlation amplitude on the
smallest scales (the significance of the offset is 2-$\sigma$).  A
tentative interpretation of this finding would be the expected higher
clustering bias of sub-mm galaxies.  This is consistent with
suggestions of clustering of sub-mm galaxies in redshift-space (Blain
et al.\ 2004).  We note that differences in the redshift distributions
of the sub-mm and optical sources will affect this comparison: the ACS
catalogue contains more objects at the lower end of the redshift range
analyzed.  However, for these low-redshift pairs a given angular scale
corresponds to a smaller physical scale, which will boost the
correlation amplitude.  Therefore, the 2-$\sigma$ significance of the
discrepancy between the cross-correlation and auto-correlation
functions in Fig.~\ref{figclus} is conservative.

There are several ways in which one might try to refine this
procedure, including: changing the weighting scheme for SCUBA galaxies
which have spectroscopic redshifts; adapting the weight of each pair
depending on the quality of the photometric redshift(s); using cuts on
colour as well as magnitude; etc.  We have not performed an exhaustive
investigation of these issues, since what is really required to
achieve definitive measurements is larger sub-mm surveys, as discussed
in Section \ref{secfutsurv}.

\section{Cross-correlation modelling}
\label{secmod}

\subsection{Cosmic magnification prediction}
\label{secmagmod}

In Section \ref{secmagmeas} we failed to detect any evidence for
cross-correlation between our sub-mm and optical data-sets of the form
which might be induced by cosmic magnification.  We now compare this
result to theoretical predictions for the size of this effect.

The magnitude of the cross-correlation amplitude induced by cosmic
magnification is determined in part by the slope of the differential
number-counts for the background population: $w_{\rm cross} \propto
(\beta-1)$, where the number of sources with fluxes between $S$ and
$S+dS$ is given by $S^{-\beta} \, dS$.  The number-counts slope for
bright sub-mm sources is known to be exceptionally steep, $\beta
\simeq 3$, increasing the effect of cosmic magnification relative to
populations with more shallow counts (such as quasars, where the
maximum slope $\beta \simeq 2$ is only reached for the very brightest
sources, Scranton et al.\ 2005).  For fainter sub-mm galaxies the
effective value of $\beta$ becomes smaller, in accordance with
equation \ref{eqfluxdist}.

The cross-correlation function due to cosmic magnification can be
predicted using simple models (e.g.\ Moessner \& Jain 1998).  If we
assume that the source and lens populations are at fixed redshifts
$z_{\rm s}$ and $z_{\rm l}$, then
\begin{equation}
w_{\rm sl}(\theta) = \frac{3 \Omega_{\rm m} (\beta-1)}{2\pi L_{\rm
    H}^2} \frac{b_{\rm l} g_{\rm ls}}{a(z_{\rm l})} \int k P(k,z_{\rm
  l}) J_0(k x(z_{\rm l}) \theta) dk ,
\label{eqmagmod}
\end{equation}
where $\Omega_{\rm m}$ is the present-day matter density relative to
the critical density, $L_{\rm H} \equiv c/H_0$ is the Hubble length in
units of $h^{-1}$ Mpc, $b_{\rm l}$ is the linear biassing factor for
the lenses, $g_{\rm ls} = x(z_{\rm l})[x(z_{\rm s}) - x(z_{\rm
    l})]/x(z_{\rm s})$ is the geometrical factor for lensing, $x(z)$
is the co-moving angular diameter distance for a flat Universe, $a(z)$
is the usual cosmological scale factor, $P(k,z)$ is the matter power
spectrum at redshift $z$ (including non-linear clustering), and
$J_0(u)$ is the zeroth-order Bessel function.  In equation
\ref{eqmagmod} the units of $x(z)$, $k$ and $P(k,z)$ are $h^{-1}$Mpc,
$h\,{\rm Mpc}^{-1}$ and $h^{-3}\,{\rm Mpc}^3$, respectively.

\begin{figure}
\center
\epsfig{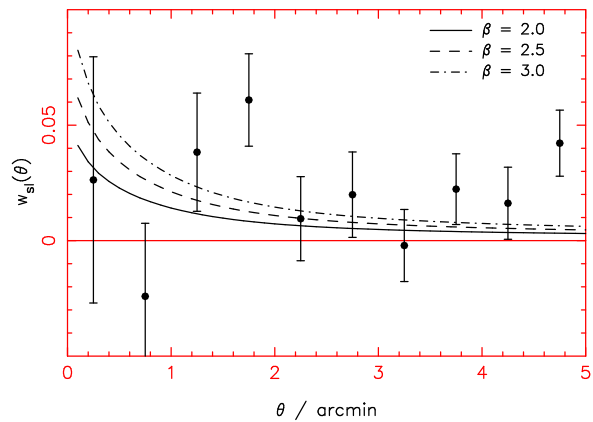}
\caption{Model cross-correlation function for cosmic magnification of
  a background source population at $z_{\rm s} = 2$ by a foreground
  lens population at $z_{\rm l} = 0.5$.  Different curves correspond
  to different slopes $\beta$ of the number-counts relation for the
  background population.  The plotted data points are a reproduction
  of our measurement from Fig.~\ref{figcosmag}.}
\label{figmagmod}
\end{figure}

We considered a foreground lens population with $b_{\rm l} = 1$ at
$z_{\rm l} = 0.5$ and a background source population at $z_{\rm s} =
2$.  We assumed a spatially-flat cosmology with cosmological
parameters $\Omega_{\rm m} = 0.3$ and $H_0 = 70$ km s$^{-1}$
Mpc$^{-1}$, and generated a non-linear power spectrum using the
prescription of Peacock \& Dodds (1994), using a linear power spectrum
produced from the fitting formulae of Eisenstein \& Hu (1998), with
baryon fraction $\Omega_{\rm b}/\Omega_{\rm m} = 0.15$, spectral index
$n_{\rm s} = 1$ and zero-redshift normalization $\sigma_8 = 0.8$
(scaled to redshift $z_{\rm l}$ using the linear growth factor of
Carroll, Press \& Turner 1992).  In Fig.~\ref{figmagmod} we plot the
resulting cross-correlation function of equation \ref{eqmagmod} for
number-count slopes between $\beta = 2.0$ and $\beta = 3.0$.
Realistically, the redshift distributions of the background and
foreground populations will be broader than the infinitesimal shells
we have assumed; this will lower the amplitude of the
cross-correlation, owing to the geometrical factor $g_{\rm ls}$ in
equation \ref{eqmagmod}.

As can be seen, the predictions of Fig.~\ref{figmagmod} provide an
entirely acceptable fit to the null measurement of cosmic
magnification plotted in Fig.~\ref{figcosmag} (which is reproduced in
Fig.~\ref{figmagmod} for comparison).  However, our results appear to
disagree somewhat with those of Almaini et al.\ (2005, Figures 1 and
2), who observed a significantly higher cross-correlation amplitude
between sub-mm and optically-selected galaxies, with their favoured
explanation being lensing magnification.  The signal-to-noise of the
measurements is low, but we find it difficult to reconcile these
results, corresponding to $w_{\rm sl} \simeq 0.2$ on small scales
$\theta \simeq 30\arcsec$, with our lensing model -- unless the fields
studied happen to contain a highly dis-proportionate concentration of
massive lenses.

As an alternative and cruder model, we can estimate the quantity of
lenses required to generate a given cross-correlation signal by
assuming a simple distribution of singular isothermal spheres.
Denoting the lensing magnification factor as $\mu$, the enhancement in
surface density of background sources is given by $\mu^{\beta-1}$,
where $\beta$ is the slope of the differential number counts
distribution (as defined above).  For a background source at (lensed)
angular separation $\theta$ from an isothermal lens with Einstein
radius $\theta_{\rm E}$,
\begin{equation}
\mu = \left( 1 - \frac{\theta_{\rm E}}{\theta} \right)^{-1}
\label{eqlens}
\end{equation}
(Bartelmann \& Schneider 2001, equation 3.19).  Assuming $\beta = 3$
(which is appropriate for the brightest SCUBA galaxies), a correlation
function $w(0.5\arcmin) \simeq 0.2$ is generated if sources have an
average magnification such that $w \simeq \mu^{\beta-1} - 1$ or $\mu
\simeq 1.1$.  Using equation \ref{eqlens}, the average source must be
at angular separation $\theta = 0.5$ arcmin from a lens with Einstein
radius $\theta_{\rm E} \simeq 2.6\arcsec$.  The Einstein radius of an
isothermal sphere with 1D velocity disperson $\sigma_v$ is given by
\begin{equation}
\theta_{\rm E} = 4 \pi \left( \frac{\sigma_v}{c} \right)^2 \left[
  \frac{x(z_{\rm s}) - x(z_{\rm l})}{x(z_{\rm s})} \right]
\end{equation}
(Bartelmann \& Schneider 2001, equation 3.17).  Substituting the
approximate value $0.5$ for the geometrical factor, we obtain
$\theta_{\rm E} \simeq (0.6\arcsec) (\sigma_v/200 {\rm \, km \,
  s}^{-1})^2$.  Hence in this simple model, lenses of velocity
dispersion $\sigma_v \simeq 420$ km s$^{-1}$ (i.e., galaxy groups)
must be responsible for the lensing magnification.  Given that our
background sources must be located on average $0.5$ arcmin from these
lenses, we estimate that a lens surface density $\simeq 1$
arcmin$^{-2}$ is required.  This is consistent with the density of
optical sources at the relevant magnitude limit, but significantly
exceeds the expected density of groups (i.e., $\simeq 0.02$
arcmin$^{-2}$, e.g.\ Yan et al.\ 2004).  We conclude that the fields
observed by Almaini et al.\ (2005) would have to be unusual areas of
sky in order to generate an angular correlation function
$w(0.5\arcmin) \simeq 0.2$ by lensing magnification.  We note that if
$w(0.5\arcmin) = 0.04$ (Fig.~\ref{figmagmod}), then the equivalent
lens velocity dispersion is $\sigma_v \simeq 200$ km s$^{-1}$, more
typical of individual galaxies with a surface density $\simeq 1$
arcmin$^{-2}$ in the appropriate redshift range.

As suggested by Almaini et al., an alternative explanation for their
observed cross-correlations is that a higher-than-expected fraction of
the sub-mm galaxies studied were located at relatively low redshifts,
$z \la 1$, and that the signal was generated by galaxy clustering
rather than by lensing magnification, as discussed in Section
\ref{secmodclus} below.  However, only $\sim 10\%$ of our sample of
sub-mm sources in GOODS-N lies at $z < 1$, and we note that, in
surveys with large redshift depth, it is difficult to generate any
significant angular cross-correlation amplitude from galaxy clustering
without photometric redshift information to restrict the redshift
slices compared.

\subsection{Galaxy clustering prediction}
\label{secmodclus}

We now estimate the amplitude of angular cross-correlation resulting
from the mutual clustering of sub-mm and optically-selected galaxies
with similar photometric redshifts.  Both populations trace the
underlying distribution of dark matter, which (for the purposes of
this calculation) we will assume is clustered in accordance with a
power-law spatial auto-correlation function, $\xi(r) =
(r/r_0)^{-\gamma}$, where $r$ is a co-moving separation, $r_0$ is the
co-moving `clustering length' at the redshift in question, and the
slope $\gamma = 1.8$.  We will further assume that the two galaxy
populations are linearly biased with respect to the dark matter
fluctuations, possessing clustering lengths $r_1$ and $r_2$.  In this
case, using the definition of bias, the spatial cross-correlation
function $\xi_{\rm cross}$ is derived by replacing $r_0$ with $(r_1
r_2)^{1/2}$ in the formula for $\xi(r)$.  The spatial
cross-correlation function, $\xi_{\rm cross}(r)$, must then be
projected onto an angular cross-correlation function, $w_{\rm
  cross}(\theta)$.

We consider the two cases of sub-mm galaxies with spectroscopic
redshifts and with only photometric redshifts.  These are compared
with an optical photo-$z$ catalogue.  In the first case, the
contribution from each sub-mm galaxy (at redshift $z = z_0$) to the
angular cross-correlation at angle $\theta$ can be determined by
integrating $\xi_{\rm cross}$ along the line-of-sight at angular
separation $\theta$, weighted by the redshift probability distribution
$p(z)$ of the optically-selected sources (normalized such that $\int
p(z) \, dz = 1$), i.e.
\begin{eqnarray}
w_{\rm cross} \hspace{-2mm} &=& \hspace{-2mm} \int p(z) \, \xi_{\rm
  cross}(\theta,z) \, dz \nonumber \\ &\simeq& \hspace{-2mm} (r_1
r_2)^{\gamma/2} \int p(z) \, [(x_0 \theta)^2 + (x -
  x_0)^2]^{-\gamma/2} dz ,
\label{eqclus1}
\end{eqnarray}
where $x(z)$ is the co-moving radial co-ordinate and $x_0 \equiv
x(z_0)$.  The redshift distribution $p(z)$ is the error distribution
of the photometric redshifts for optical galaxies with best-fitting
redshifts near $z = z_0$.  We will assume that this function is a
Gaussian distribution with mean $z_0$ and standard deviation
$\sigma_z$.  As the width of this function is much larger than the
clustering length, a good approximation for equation \ref{eqclus1} is
\begin{equation}
w_{\rm cross} = C_\gamma (r_1 r_2)^{\gamma/2} p(z_0) \left(
\frac{dx}{dz}(z_0) \right)^{-1} x(z_0)^{1-\gamma} \theta^{1-\gamma} ,
\label{eqclus2}
\end{equation}
where $C_\gamma = \Gamma(\frac{1}{2}) \, \Gamma{(\frac{\gamma}{2} -
  \frac{1}{2}}) / \Gamma(\frac{\gamma}{2})$.

For the case where the sub-mm source redshift is only photometric, in
order to obtain the resulting angular cross-correlation we must
integrate equation \ref{eqclus2} over redshift again, weighting by a
further factor of $p(z)$.  Hence we assume (for the purposes of this
calculation) that the photometric redshift error distribution for the
sub-mm galaxies is the same as for the optical sources.  The result is
an extra damping of the cross-correlation amplitude:
\begin{equation}
w_{\rm cross} = C_\gamma (r_1 r_2)^{\gamma/2} \theta^{1-\gamma} \int
p(z)^2 \left( \frac{dx}{dz} \right)^{-1} x(z)^{1-\gamma} dz .
\label{eqclus3}
\end{equation}
Equation \ref{eqclus3} is in fact the usual Limber equation for the
projection of spatial clustering (e.g.\ Peebles 1980).

\begin{figure}
\center
\epsfig{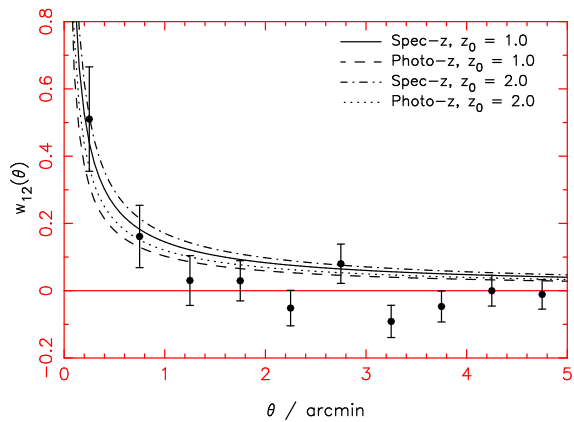}
\caption{Model cross-correlation function generated by the mutual
  clustering of sub-mm and optically-selected galaxies.  We compare
  cases where the sub-mm galaxy redshift $z_0$ is known with
  spectroscopic and photometric accuracy, for $z_0 = 1$ and $z_0 = 2$.
  We assume that the optical galaxies always have a photometric
  redshift distribution centred at $z_0$ with an r.m.s. width
  $\sigma_z = 0.2$.  The plotted data points are a reproduction of our
  measurement from Fig.~\ref{figclus}.}
\label{figclusmod}
\end{figure}

In order to evaluate these expressions, we take a photometric-redshift
error $\sigma_z = 0.2$, which is characteristic of our data
(i.e.\ $\sigma_z/(1+z) \simeq 0.1$ as noted in Section \ref{secdata}).
We assume that the co-moving clustering length of the
optically-selected galaxies is constant with redshift (Lahav et
al.\ 2002), $r_1 = 5 \, h^{-1}$ Mpc, and take an enhanced clustering
amplitude for the sub-mm galaxies, $r_2 = 7 \, h^{-1}$ Mpc (Blain et
al.\ 2004).  The predictions of equations \ref{eqclus2} and
\ref{eqclus3} for the cases of spectroscopic and photometric redshifts
for the sub-mm galaxies are displayed in Fig.~\ref{figclusmod} for
$z_0 = 1$ and $z_0 = 2$.  Finally, we make the approximation that each
sub-mm galaxy is an independent probe of the clustering, thus the
angular correlation functions determined for each sub-mm source (which
may have a spread as indicated in Fig.~\ref{figclusmod}) can simply be
averaged.  Since approximately half of the sub-mm galaxies have
spectroscopic redshifts, the final result should fall somewhere in
between the curves shown in Fig.~\ref{figclusmod}.  This model is a
reasonable fit to our observations, which are also reproduced in
Fig.~\ref{figclusmod} for comparison.

\section{Prospects for future surveys}
\label{secfutsurv}

In order to measure the cosmic magnification pattern with reasonable
accuracy (say, signal-to-noise exceeding 3 in several separation bins)
we require cross-correlation function measurements with precision
$\delta w \sim 0.002$ in bins of width $\delta \theta \sim 0.5$ arcmin
(see Fig.~\ref{figmagmod}).  The error in the correlation function is,
roughly speaking, determined by the number of galaxy pairs measured:
$\delta w \sim N_{\rm pairs}^{-1/2}$ (with the caveats discussed in
Section \ref{secest}).  If $\Sigma_{\rm opt}$ is the surface density
of optical galaxies in the appropriate redshift range, then each
sub-mm source participates in an average of $\Sigma_{\rm opt} \times 2
\pi \theta \, \delta \theta$ pairs in a bin of average separation
$\theta$.

For the current cosmic magnification analysis (Fig.~\ref{figcosmag}),
we have a surface density of optical sources $\Sigma_{\rm opt} \simeq
10$ arcmin$^{-2}$ in the appropriate redshift range ($0.2 < z < 0.8$).
Since $N_{\rm sub-mm} \simeq 30$, we recover $\delta w \simeq 0.03
\times \theta{\rm (arcmin)}^{-1/2}$ in bins of width $\delta \theta =
0.5$ arcmin, as observed in Fig.~\ref{figcosmag}.  In order to achieve
a measurement of cosmic magnification with reasonable significance we
therefore require a sub-mm survey with $\sim 100$ times as many
sources.  This should easily be achieved with surveys being planned
with the new SCUBA-2 instrument (Holland et al.\ 2003).  However, it
is worth bearing in mind that a high density of optical galaxies and
reasonably good photo-$z$ estimates will be required.

\begin{figure}
\center
\epsfig{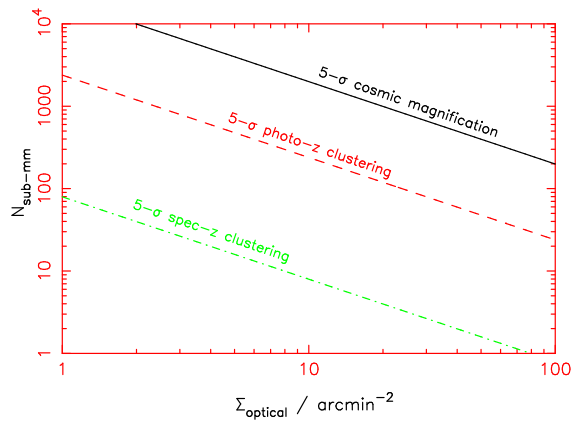}
\caption{A rough indication of the accuracy with which sub-mm surveys
  yielding $N_{\rm sub-mm}$ sources can measure cosmic magnification
  and galaxy clustering through comparison with deep optical
  catalogues with surface density $\Sigma_{\rm opt}$.  The 5-$\sigma$
  threshold for each type of analysis is defined using estimates for
  the cross-correlation function amplitude and error within an assumed
  range of measured scales.  The amplitudes are obtained from the
  correlation function models developed in Section \ref{secmod}.  The
  errors are derived from the number of object pairs within the
  separation bin, assuming a Poisson error $\delta w = N_{\rm
    pairs}^{-1/2}$.  For the cosmic magnification analysis, we assume
  $w(\theta) = 0.02$ at a scale $\theta = 1$ arcmin
  (Fig.~\ref{figmagmod}) and consider a bin of width $\delta \theta =
  0.5$ arcmin.  For the photo-$z$ clustering measurement we use the
  same separation bin, but assume $w(1 \, {\rm arcmin}) = 0.1$
  (Fig.~\ref{figclusmod}) and divide the catalogues into three
  independent photo-$z$ bins.  For the estimate where all objects have
  spectroscopic redshifts, we spread the optical sources over a
  redshift range $1 < z < 2$ and consider the measurement of a spatial
  correlation function $\xi = 1$ at a scale $5 \, h^{-1}$ Mpc in a bin
  of width $1 \, h^{-1}$ Mpc.}
\label{figfut}
\end{figure}

The cross-correlation resulting from the mutual clustering of sub-mm
and optically-selected galaxies should be easier to detect, owing to
its larger amplitude (Fig.~\ref{figclusmod}).  Let us assume optical
data of equivalent quality to that used in this study.  The surface
density of optical galaxies with `high-quality' photometric redshifts
in a redshift range overlapping with the sub-mm sources is again
$\Sigma_{\rm opt} \simeq 10$ arcmin$^{-2}$, which we divide into 3
independent photo-$z$ bins.  Since $N_{\rm sub-mm} \simeq 15$ in the
region of overlap, we recover $\delta w \simeq 0.08 \times \theta{\rm
  (arcmin)}^{-1/2}$ in bins of width $\delta \theta = 0.5$ arcmin.
However, the expected amplitude of the cross-correlation due to
clustering (Fig.~\ref{figclusmod}) is significantly larger than that
due to cosmic magnification (Fig.~\ref{figmagmod}): $w_{\rm cross}
\simeq 0.1 - 0.2$ at $\theta = 1$ arcmin.  We therefore conclude that
a sub-mm survey with $\sim 10$ times as many sources as used in the
current study should suffice to measure this signal with reasonable
accuracy, assuming an appropriate quantity of optical follow-up.  This
should be achievable for the on-going SHADES project (Mortier et
al.\ 2005).  Alternatively, we note that the clustering amplitude
could be accurately measured using the current catalogues if {\it
  spectroscopic} redshifts were available for both the optical and
sub-mm sources.  Figure \ref{figfut} illustrates the accuracy of
measurement achievable for each type of analysis through the
comparison of sub-mm and optical catalogues.

\section{Summary}

We have investigated the cross-correlation between sub-mm and optical
sources in the GOODS-N survey region, using photometric redshift
information.  We find that:
\begin{itemize}
\item Comparing high-redshift $(z > 1)$ SCUBA sources with
  low-redshift $(0.2 < z < 0.8)$ optical galaxies, we can detect no
  evidence for cross-correlation due to cosmic magnification.  We
  attribute previous reported detections to either: (1) a difference
  in the correlation function estimator employed; (2) analysis of a
  field that happens to contain a highly dis-proportionate
  concentration of massive lenses; or (3) a higher-than-expected
  fraction of sub-mm galaxies residing at relatively low redshifts $z
  \la 1$.  Based on calculations of the expected amplitude of the
  lensing magnification signal in the standard cosmology, the sub-mm
  data-set must be increased in size by a factor $\simeq 100$ to
  secure a significant measurement.
\item Comparing optical and sub-mm sources in identical photometric
  redshift slices, we detect evidence for a cross-correlation due to
  galaxy clustering (with a significance level of $3.5$-$\sigma$).
  The sub-mm sources appear to possess a higher bias factor than the
  optical galaxies (with a significance of 2-$\sigma$).  This
  observation, if confirmed by larger surveys, would support the
  hypothesis that sub-mm sources form in relatively dense environments
  in the high-redshift Universe.
\end{itemize}

One of the primary goals of the SHADES project (Mortier et al.\ 2005)
is to measure the clustering properties of sub-mm galaxies via an
auto-correlation function analysis.  We note that the
cross-correlation with optically-selected galaxies could also provide
valuable information (owing to the higher surface density of optical
sources), provided that the optical data are of sufficient quality.
Cross-correlation in different redshift slices, in order to measure
the lensing magnification, is an independent effect which should add
an extra structure formation diagnostic to future sub-mm surveys, such
as those that will be carried out with SCUBA-2.  In terms of structure
formation models, the auto- and cross-correlation functions have a
different dependence on the halo occupation distribution, as well as
on redshift and other source properties.  Hence, an investigation of
cross-correlation in future ambitious sub-mm surveys holds the promise
of unravelling details of galaxy formation and bias within massive
haloes.

\section*{Acknowledgments}

CB acknowledges support from the Izaak Walton Killam Memorial Fund for
Advanced Studies and from the Canadian Institute for Theoretical
Astrophysics National Fellowship programme.  This research has been
supported by the Natural Sciences and Engineering Research Council of
Canada.  We thank Ludo van Waerbeke for useful discussions.

\end{document}